\documentclass[twocolumn,amsmath,amssymb,prl,aps]{revtex4}
\input psfig
\begin{document}

\title{Does the Sun Shine by $\mathbf{pp}$ or CNO Fusion Reactions?}
\author{John N. Bahcall\thanks{E-mail: jnb@ias.edu}}
\affiliation{School of Natural Sciences, Institute for Advanced Study, Princeton, New Jersey 08540}
\author{M. C. Gonzalez-Garcia
\thanks{E-mail: concepcion.gonzalez-garcia@cern.ch}}
\affiliation{Theory Division, CERN, CH-1211, Geneva 23, Switzerland,
and Y.I.T.P., SUNY at Stony Brook, Stony Brook, NY 11794-3840, USA,
and IFIC, Universitat de Val\`encia - C.S.I.C., Apt 22085, 46071
Val\`encia, Spain}
\author{Carlos Pe\~na-Garay\thanks{E-mail: penya@ias.edu}}
\affiliation{School of Natural Sciences, Institute for Advanced Study,
Princeton, NJ 08540, USA}

\begin{abstract}
We show that solar neutrino experiments set an upper limit of 7.8\% (7.3\% including the recent KamLAND
measurements) to the fraction of energy that the Sun produces via the CNO fusion cycle, which is an order
of magnitude improvement upon the previous limit. New experiments are required to detect CNO neutrinos
corresponding to the 1.5\% of the solar luminosity that the standard solar model predicts is generated by
the CNO cycle.

\end{abstract}

\pacs{Valid PACS appear here}
\maketitle

In 1939, Hans Bethe described in an epochal paper~\cite{bethe39} two nuclear fusion mechanisms by which
main sequence stars like the Sun could produce the energy corresponding to their observed luminosities.
The two mechanisms have become known as the $p-p$ chain and the CNO cycle~\cite{book}. For both the $p-p$
chain and the CNO cycle the basic energy source is the burning of four protons to form an alpha
particle,two positrons, and two neutrinos. Thus
\begin{equation}
4p ~\rightarrow~ ^4{\rm He}~+2e^+ ~+~2\nu_e ~+~ \leq 25 {\rm ~MeV~(to~the~star)}.\label{eq:fourprotons}
\end{equation}
In the $p-p$ chain, fusion reactions among elements lighter than $A = 8$ produce a characteristic set of
neutrino fluxes, whose spectral energy shapes are known but whose fluxes must be calculated with a
detailed solar model.  In the CNO chain, with $^{12}$C as a catalyst, $^{13}$N and $^{15}$O beta decays
are the primary source of neutrinos.

The first sentence in Bethe's paper reads: ``It is shown that the most important source of energy in
ordinary stars is the reactions of carbon and nitrogen with protons." Bethe's conclusion about the
dominant role of the CNO cycle relied upon a crude model of the Sun. Over the next two and a half
decades, the results of increasingly more accurate laboratory measurements of nuclear fusion reactions
and more detailed solar model calculations led to the theoretical inference that the Sun shines primarily
by the $p-p$ chain rather than the CNO cycle. Currently, solar model calculations imply~\cite{bp00} that
98.5\% of the solar luminosity is provided by the $p-p$ chain and only $1.5$\% is provided by CNO
reactions.

In recent years, there have been many analyses of solar neutrino oscillations, essentially all of which
assumed that the CNO neutrino fluxes were equal to their predicted standard solar model values (see
Refs.~\cite{postsno02,snodaynight,bargernc} and references cited therein). However, from the earliest
days of solar neutrino research, a primary goal of the field was to test the solar model prediction that
the Sun shines by the $p-p$ chain and not by the CNO cycle~\cite{bahcall69}. This goal has largely been
ignored in the last decade or so as solar neutrino experiments concentrated on the more accessible,
higher-energy $^8$B neutrinos. In this paper, we return to the question of how well we can measure, or
set an upper limit to, the CNO neutrino fluxes.

Unfortunately, the standard solar model prediction for the CNO fluxes is difficult to test. Radiochemical
experiments with chlorine~\cite{chlorine} and gallium~\cite{sage,gallex,gno} do not measure the energy of
the neutrinos detected; they measure the rate of neutrino induced events above a fixed energy threshold.
The neutrino-electron scattering experiments, Kamiokande~\cite{kamiokande} and
Super-Kamiokande~\cite{superk}, provide information about neutrinos but only those that have energies
well above the maximum energies of the $^{13}$N ($E_{\rm max} = 1.2$ MeV) and $^{15}$O ($E_{\rm max} =
1.7$ MeV) neutrinos. The radiochemical experiments are only sensitive to electron type neutrinos, and the
neutrino-electron scattering experiments are primarily sensitive to electron type neutrinos. The heavy
water experiment, SNO~\cite{snodaynight,snoccnc}, measures higher energy neutrinos. The goal of uniquely
identifying CNO neutrinos is made even more difficult by the fact that neutrino oscillations can change
in an energy dependent way the probability that electron type neutrinos created in the Sun reach the
Earth as electron type neutrinos~\cite{pontecorvo,msw}.

Because of these complications, it was possible to find neutrino oscillation solutions in which 99.95 \%
of the Sun's luminosity is supplied by the CNO cycle~\cite{cno}. These `large CNO' oscillation solutions
describe well all of the measurements from the chlorine~\cite{chlorine}, SAGE~\cite{sage},
GALLEX~\cite{gallex}, and Kamiokande~\cite{kamiokande} solar neutrino experiments. Modern solar models do
not predict a large CNO contribution to the solar luminosity, but the goal is to test experimentally-not
just assume-this prediction.

In this paper, we use data from  the chlorine, SAGE, GALLEX, GNO, Super-Kamiokande, and SNO solar
neutrino experiments, and from the recent KamLAND~\cite{kamland} reactor measurements, to set an
experimental limit on the CNO contribution to the solar luminosity that is an order of magnitude more
stringent than the previous best limit~\cite{cno}. Although individual experiments do not constrain well
the CNO fluxes, a global solution to all the available neutrino data  provides a powerful upper limit. We
also discuss how well future experiments can do in detecting the CNO neutrinos.

Here is our strategy. For each value of the CNO luminosity fraction, $L_{\rm CNO}/L_\odot$, we search the
 two-component neutrino oscillation parameter space with a dense mesh corresponding to the neutrino mass difference $10^{-12} {\rm eV^2} <
\Delta m^2 < 10^{-3} {\rm eV^2}$ (721 mesh points) and mixing angles $0.0001 < \tan^2 \theta < 10$ (401
mesh points), as well as the solar neutrino fluxes (see below).  [We verify later that our approximation
of only two neutrinos does not limit the validity of the upper bound we derive. See discussion following
Eq.~(\ref{eq:maxflux}).] We calculate the global $\chi^2$ by fitting to all the available data,
\begin{equation}
\chi^2 ~=~ \chi^2_{\rm solar} ~+~ \chi^2_{\rm KamLAND}\,.   \label{eq:chisquared}
\end{equation}
We carry out a global analysis~\cite{postsno02} of the solar neutrino data letting the neutrino fluxes be
free variables and using data from 80 measurements:  44 data points from the Super-Kamiokande
zenith-angle energy distribution, 34 data points from the SNO day-night energy
spectrum~\cite{snoccnc,snodaynight}, and 2 radio-chemical rates from Cl~\cite{chlorine} and
Ga~\cite{sage,gallex,gno}. We define the $3\sigma$ upper limit for the CNO neutrino fluxes by determining
when $\chi^2 = \chi^2_{min} + 9$ after marginalizing over the oscillation parameters $\Delta m^2$ and
$\tan^2 \theta$ and over the other neutrino fluxes.

Using the data provided in Ref.~\cite{kamland}, we calculate the positron spectrum in the KamLAND
detector with the procedures described in Refs.~\cite{kamland,prekamlandus,prekamlandthem}.  In the
absence of neutrino oscillations, we find (in agreement with Ref.~\cite{kamland}) 86.8 expected neutrino
events above 2.6 MeV visible energy for the stated experimental conditions. The positron energy spectrum
that we calculate is in excellent agreement with the energy spectrum presented by the KamLAND
collaboration. Further details of our  analysis of the KamLAND and solar data  can be found in
Ref.~\cite{postkamland}.

We impose the `luminosity constraint' on the solar neutrino fluxes, i.e., we require that the sum of the
thermal energy generation rates associated with each of the solar neutrino fluxes be equal to the solar
luminosity~\cite{spiro}. The fraction of the sun's luminosity that arises from CNO reactions can be
written as~\cite{luminosity} :
\begin{equation}
\frac{L_{CNO}}{L_\odot} ~=~ \sum\limits_{i=N,O,F} (\frac{\alpha_i}{10~ {\rm MeV}}) a_i \phi_i \,,
\label{eq:cnosum}
\end{equation}
where the constant $\alpha_i$ is the energy provided to the star by nuclear fusion reactions associated
with the $i^{th}$ neutrino flux, $a_i$ is the ratio of the neutrino flux $\Phi_i {\rm (BP00)}$ of the
standard solar model to the characteristic solar photon flux defined by $L_\odot/[4\pi(A.U.)^2({\rm 10
MeV})]$, and $\phi_i$ is the ratio of the true solar neutrino flux to the neutrino fluxes predicted by
the BP00 standard solar model \cite{bp00}. Ref.~\cite{luminosity} presents a detailed derivation of
Eq.~(\ref{eq:cnosum}) and the numerical values for the coefficients $\alpha_i$ and $a_i$.

We treat as free parameters all the solar neutrino fluxes that are normally reported in solar neutrino
calculations. There are then 10 free parameters: the two oscillation parameters, $\Delta m^2$ and
$\tan^2\theta$,and the 8 neutrino fluxes, $p-p$, $pep$, $^7$Be,$^8$B, and hep (from the $p-p$ chain) and
$^{13}$N $^{15}$O, and $^{17}$F (from the CNO cycle). To speed up the calculations, we made some
approximations that we have checked do not affect the accuracy of our search. Two of the solar neutrino
fluxes, hep and $^{17}$F, are small as a result of nuclear physics considerations. In the initial search
calculations, we set hep equal to its solar model value and $^{17}{\rm F} = \left[\left(^{17}{\rm
F}/^{13}{\rm N}\right)_{\rm solar~model}\right] ^{13}{\rm N}$. We checked that our results are unchanged
if the hep solar model flux is multiplied by eight (present experimental bound from the high energy bins
at Super-Kamiokande~\cite{smy}) or if we set the $^{17}$F flux equal to the $^{17}$N flux. Also, the
ratio of the $pep$ neutrino flux to the $p-p$ neutrino flux is fixed to high accuracy because they have
the same nuclear matrix element. We have set the ratio equal to the standard solar model value and have
checked that our results are unchanged if this ratio is varied by 10\% (an enormous change). We set the
$^{13}$N flux equal to the $^{15}$O flux, which is expected in the limit that the CNO contribution to the
luminosity is dominant. We also verified that the results of our search are unchanged if we set the ratio
of $^{13}$N to $^{15}$O neutrino fluxes equal to the standard solar model value, the ratio expected if
the $p-p$ contribution is dominant. Finally, we checked several intermediate values of this ratio to see
that the upper limit we quote here is robust and valid in all cases.

We find the minimum value of $\chi^2$ for each assumed value of the CNO luminosity fraction by
marginalizing over the neutrino oscillation parameters and over the non-CNO neutrino fluxes. We performed
the calculations in two stages: first using only the solar neutrino data and second using both the solar
neutrino and the KamLAND data. We carried out calculations for oscillations to purely active neutrinos,
to purely sterile neutrinos, and to active-sterile admixtures as described in Ref.~\cite{four} (see also
last reference in Ref.~\cite{prekamlandus}). We  considered sterile admixtures that range from the
maximum allowed by the recent KamLAND data, $\sin^2\eta = 13$\%~\cite{postkamland}, to  25\%, 50\%, 75\%,
as well as the extremes of 0\% and 100\%. For all values of the CNO luminosity fraction,  the minimum
$\chi^2$ was, as expected, achieved for purely active oscillations.

\begin{figure}[!t]
\centerline{\psfig{figure=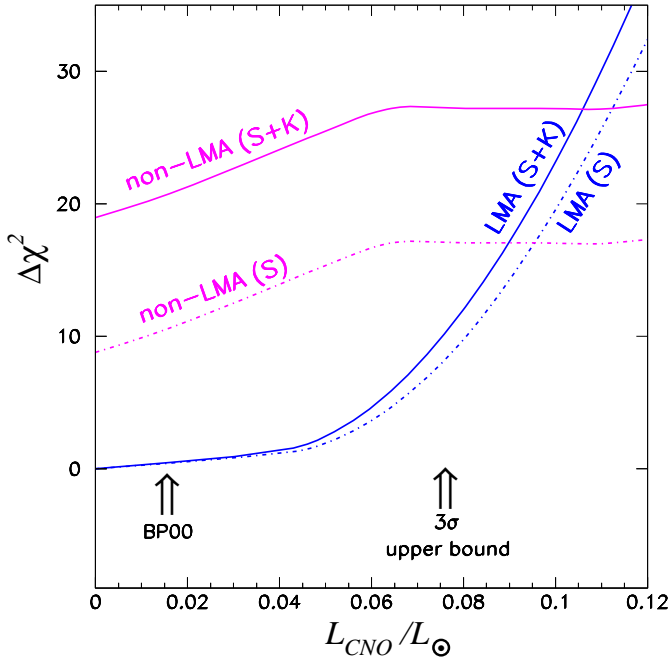,width=3.5in}} \caption[]{Experimental bound on $L_{\rm CNO}/L_{\odot}$.
The figure shows how the $\chi^2$ fit worsens as one increases the assumed fraction of the Sun's
luminosity that arises from CNO reactions. The dotted lines were computed using just solar neutrino data;
the solid lines use both solar neutrino experiments and the KamLAND results.The curves labeled LMA were
calculated for the favored large mixing angle MSW solution; the curves labeled non-LMA were calculated
for the best fit of the LOW, SMA, vacuum, and sterile oscillation solutions.  The arrows indicate the
predicted 1.5\% CNO luminosity from the standard solar model~\cite{bp00} and the $\sim 8$ \% (see
Eq.~\ref{eq:cnolimit}) $3\sigma$ upper bound \hbox{(1 dof)} allowed by the chlorine, gallium,
Super-Kamiokande, and SNO solar neutrino data~\cite{snodaynight,chlorine,sage,gallex,superk,snoccnc} and
the KamLAND reactor data~\cite{kamland}. \label{fig:cnol}}
\end{figure}

Figure~\ref{fig:cnol} summarizes our main results. The figure shows $\Delta \chi^2$ as a function of the
CNO luminosity fraction when only solar neutrino data are used (denoted by dotted curves) and when solar
and KamLAND data are used (denoted by solid curves). The minimum value of $\chi^2$, relative to which
$\Delta \chi^2$ is measured, is reached in both cases for a zero value of the CNO flux. However, as is
apparent from Figure~\ref{fig:cnol}, the global $\chi^2$ is essentially flat for all values of $L_{\rm
CNO}/L_\odot < 5$\%. Current experiments are not sensitive to CNO neutrino fluxes that correspond to less
than 5\% of the solar luminosity. For example, the $\Delta \chi^2$ is only 0.5 between the best fit at
$L_{\rm CNO}/L_\odot = 0.0$ and the standard solar model value of 1.5\%.

In all cases, the best fit is a LMA (large mixing angle) MSW solution~\cite{msw}. In principle, new
solutions might have been found by allowing $L_{\rm CNO}/L_\odot $ to be a free parameter. In practice,
no preferred new solutions are found for any value of $L_{\rm CNO}/L_\odot $.

For completeness, we explored the entire oscillation parameter space specified just preceding
Eq.~(\ref{eq:chisquared}), including the parameter ranges of all previously recognized oscillation
solutions~\cite{postsno02,snodaynight,bargernc}. The best-fit results for these searches for other
oscillation cases are labeled `non-LMA' in Figure~\ref{fig:cnol}. For $L_{\rm CNO}/L_\odot$ less than
about 5\%, the best-fit non-LMA solution is the well-known LOW solution. For the larger values of $L_{\rm
CNO}/L_\odot$ shown in Figure~\ref{fig:cnol}, the best-fit LMA solution is a vacuum solution with $\Delta
m^2 = 7.9\times 10^{-11} {\rm eV^2}$ and $\tan^2 \theta = 0.22$. We conclude, using both solar neutrino
and KamLAND experimental data (using just solar neutrino data), that
\begin{equation}
\frac{L_{\rm CNO}}{L_\odot} ~<~ 7.3 \% ~(7.8 \%)~~{\rm at ~3\sigma} \,. \label{eq:cnolimit}
\end{equation}
The recent KamLAND measurements reduce the upper limit by 0.5\%.

 The order of magnitude improvement
between the previous limit of 99.95\%\cite{cno} and the present limit, Eq.~(\ref{eq:cnolimit}), is due to
the Super-Kamiokande and SNO measurements. The earlier large CNO oscillation solution was confined to
small mixing angles, SMA, which cannot fit simultaneously the flat recoil energy spectrum measured by
Super-Kamiokande~\cite{superk} and the total event rates measured by Super-Kamiokande and
SNO~\cite{superk,snoccnc}.

The maximum  CNO neutrino flux allowed by the existing experiments is
\begin{equation}
\phi_{\rm CNO,~max} < 3.41\times10^{10}{\rm \, cm^{-2}s^{-1}}\left(L_{\rm CNO}/L_\odot\right)\,,
\label{eq:maxflux}
\end{equation}
where $\phi_{\rm CNO,~max}= \phi({\rm ^{13}N})_{\rm max} = \phi({\rm ^{15}O})_{\rm max}$.

We have verified that the upper limits given in Eq.~(\ref{eq:cnolimit}) and Eq.~(\ref{eq:maxflux}) are
not affected by the approximation of assuming that there is just one mass scale (i. e., two neutrinos).
We repeated the analysis assuming the standard three-neutrino mixing scenario invoked to explain both
solar and atmospheric data and  assumed values for $\theta_{13}$ values below the CHOOZ
bound~\cite{chooz}, $\tan^2(\theta_{13})=0.0$, $0.03$ and $0.06$ (the CHOOZ bound at
3$\sigma$~\cite{concha13}). The minimum $\chi^2$ was, as expected, achieved for
$\tan^2(\theta_{13})=0.0$.

New solar neutrino experiments are required to measure the CNO contribution to the solar luminosity.

How much can a future $^7$Be neutrino-electron scattering experiment, BOREXINO~\cite{borexino} or
KamLAND~\cite{kamland}, improve the limit given in Eq.~(\ref{eq:cnolimit})? We find an   approximate
answer to this question by computing a global $\chi^2$ including the existing solar neutrino data, the
KamLAND reactor data,  and a simulated BOREXINO rate measurement (simulations guided by
Ref.~\cite{borexino}). We assume that the BOREXINO rate will be consistent with the predicted best fit
point from the solar plus KamLAND global fit with a total error of 10\% (5\%)  for the rate measurement.
If these assumptions are valid, one will be able to either measure $L_{\rm CNO}/L_\odot$ or conclude that
$L_{\rm CNO}/L_\odot < 5.6\% \,(4.9)$\%.

In order to measure the CNO contribution at the 1.5\% level predicted by the standard solar model, one
must be able to distinguish the continuum $^{13}$N and $^{15}$O neutrinos from the $^7$Be and $pep$
neutrino lines, as well as from all the sources of background. The appropriate analyses of proposed low
energy neutrino-electron scattering detectors have not yet been carried out, so one cannot say for sure
whether or not this will be possible. But, it seems very difficult. The energy  resolution required to
measure the energy of the CNO neutrinos and determine their flux, may, however, be within the reach of
low-energy CC experiments~\cite{lowe}.

The solar model predictions for CNO neutrino fluxes are not precise  because the CNO fusion reactions are
not as well studied as the $p-p$ reactions~\cite{adelberger} and because the Coulomb barrier is higher
for the CNO reactions, implying a greater sensitivity to details of the solar model. For the standard
solar model CNO neutrino fluxes, the $1\sigma$ errors vary between 17\% and 25\%~\cite{bp00}. A
measurement of the CNO  neutrino fluxes would constitute a stringent test of the theory of stellar
evolution and provide unique information about the solar interior.

 \acknowledgments JNB and CPG acknowledge support from NSF grant No. PHY0070928. MCG-G
is supported by the European Union Grant No HPMF-CT-2000-00516, by Spanish Grants No PB98-0693 and
CTIDIB/2002/24.

\end{document}